\documentclass[twocolumn,aps,prb,superscriptaddress,floatfix]{revtex4-1}
\usepackage{xcolor}
\usepackage{graphicx}
\usepackage{dcolumn}
\usepackage{bm}
\usepackage{amsmath}
\usepackage[T1]{fontenc}
\usepackage[unicode=true,pdfusetitle,
 bookmarks=true,bookmarksnumbered=false,bookmarksopen=false,
 breaklinks=false,pdfborder={0 0 1},backref=false,colorlinks=false]
 {hyperref}
\hypersetup{
 colorlinks,linkcolor=blue,citecolor=blue,urlcolor=blue}

\newcommand{\fig}[1]{FIG. #1}
\newcommand{\subfig}[2]{\fig{#1} (#2)}

\begin{document}

\preprint{APS/123-QED}

\title{Spin-orbit torque emerging from orbital textures in centrosymmetric materials}

\author{Luis M. Canonico}
\affiliation{Catalan Institute of Nanoscience and Nanotechnology (ICN2), CSIC and BIST, Campus UAB, Bellaterra, 08193 Barcelona, Spain}
\author{Jose H. Garcia}
\email{josehugo.garcia@icn2.cat}
\affiliation{Catalan Institute of Nanoscience and Nanotechnology (ICN2), CSIC and BIST, Campus UAB, Bellaterra, 08193 Barcelona, Spain}
\altaffiliation{Corresponding author. Email: josehugo.garcia@icn2.cat}
\author{Stephan Roche}
\affiliation{Catalan Institute of Nanoscience and Nanotechnology (ICN2), CSIC and BIST, Campus UAB, Bellaterra, 08193 Barcelona, Spain}
\affiliation{ICREA, Instituci\'o Catalana de Recerca i Estudis Avançats, 08070 Barcelona, Spain}

\date{\today}

\begin{abstract}
We unveil a hitherto concealed spin-orbit torque mechanism driven by orbital degrees of freedom in centrosymmetric two-dimensional transition metal dichalcogenides (focusing on PtSe${}_2$). Using first-principles simulations, tight-binding models and large-scale quantum transport calculations, we show that such a mechanism fundamentally stems from a spatial localization of orbital textures at opposite sides of the material, which imprints their symmetries onto spin-orbit coupling effects, further producing efficient and tunable spin-orbit torque. Our study suggests that orbital-spin entanglement at play in centrosymmetric materials can be harnessed as a resource for outperforming conventional spin-orbit torques generated by the Rashba-type effects. 
\end{abstract}

\maketitle

Spin-orbit torque (SOT) is a central mechanism mediated by charge-to-spin conversion that enables efficient manipulation of magnetization in spintronic devices \cite{gambardella2011current,manchon2019current,shao2021roadmap}. SOT is usually generated by nonequilibrium manifestations of the spin-orbit coupling (SOC) such the spin-Hall and Rashba-Edelstein effects which accumulate intrinsic angular momentum source that exerts some torque on nearby magnets. These mechanisms are well described in \emph{noncentrosymmetric} systems \cite{bihlmayer2022rashba} and are fundamentally related to spin-momentum locking in reciprocal space. Other mechanisms have been discussed for low-dimensional systems, including the contribution of the Berry curvature \cite{kurebayashi2014antidamping} or skew scattering \cite{sousa2020skew} effects, but they require inversion symmetry breaking, so to date, all known SOT mechanisms exclude the large family of \emph{centrosymmetric} materials. 

Recently, the possibility of unconventional spatially-dependent spin-momentum locking properties has been discussed for a particular family of \emph{centrosymmetric} materials, namely the 1T-phase  transition metal dichalcogenides (TMDs). In these materials, the emergence of hidden spin textures is dictated by the presence of dipolar fields within the crystal structure \cite{zhang2014hidden,yuan2019uncovering}, as confirmed by ARPES experiments \cite{PtSe2Spin-Layer,clark2022hidden}. On the other hand, the electrical manipulation of the orbital degrees of freedom can give rise to the orbital-Hall effect, where a longitudinal current drives a sizable transverse flow of orbital angular momentum even in systems with negligible SOC \cite{OHEBernevig,OrbitalTexture,OHEmetals,OHE_Bhowal_1,OHE_Bhowal-Vignale,Us2,Us3,pezo2022orbital,choi2023observation}. Such orbital effects have been shown to also signal the presence of high-order topological phases \cite{costa2022connecting}, pointing to novel paradigms in ``orbitronics". Importantly, the role of the orbital degrees of freedom and intertwined spin response was recognized in a more general description of the Rashba SOC \cite{OrbitalRashba1,OrbitalRashbaDetect,bihlmayer2022rashba,kim2022orbital} and the driving or enhancement of SOT \cite{Orbital-torque-2,Orbital-torque-1,orbital-torque-magnetic-bilayers-EXP}. However, a fundamental orbital-based description of complex spin phenomena such as the hidden spin textures and their ultimate nature in generating novel SOT effects, remains largely unexplored.

In this letter, we predict a novel SOT mechanism in \emph{centrosymmetric} materials (such as monolayer PtSe${}_2$) whose origin stems from intrinsic orbital textures of the electronic states entangled with their spin features. The hybridization between orbitals with different symmetries is the mechanism behind the emergence of this unconventional SOT, which can be further tailored using external electric fields. The theoretical results are obtained using realistic tight-binding models elaborated with first-principles simulations and implemented in large-scale (real-space) quantum transport simulations. These discovered new SOT components in \emph{centrosymmetric} structures enlarge the portfolio of available materials for designing ultracompact 2D materials-based spintronic architectures in the context of non-volatile memory technologies \cite{yang2022two}. 

\begin{figure*}[t]
\includegraphics[width=0.85\linewidth]{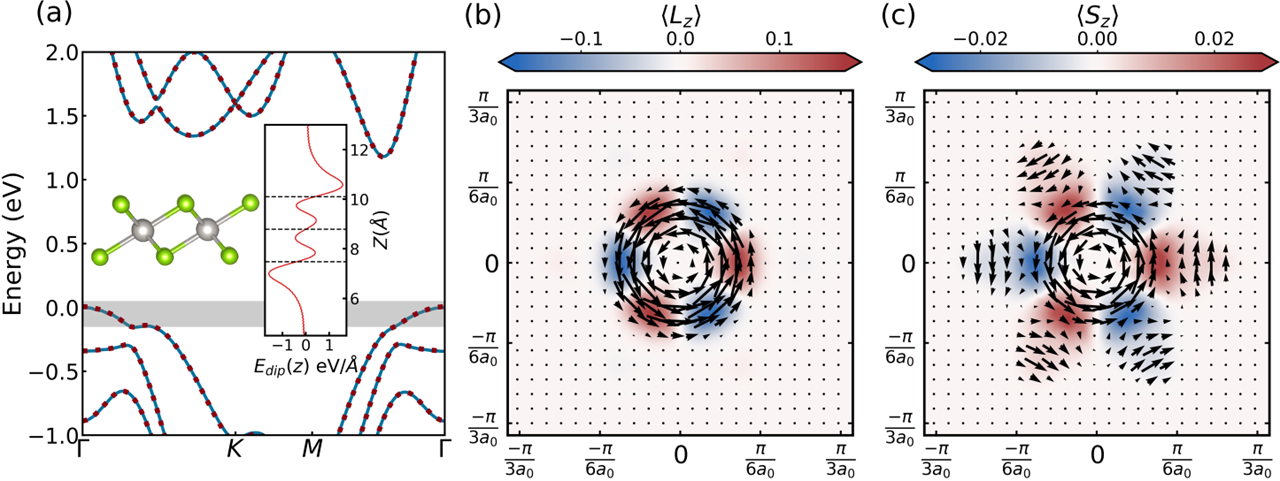}
\caption{ (a) Bandstructure of free-standing 
 monolayer PtSe$_2$ computed using fully relativistic band structure DFT (red dotted lines) and PAO Hamiltonian (blue solid lines). Inset: Side view of PtSe${}_2$ monolayer (left) where its height is chosen to match the $y$-axis of the dipolar electric field as a function of the $z-$direction computed from DFT (right). (b) Orbital texture of PtSe$_{2}$ projected in the top Se atoms, computed in the energy window shown in (a) (grey area), where $L_x(\bm{k})$ and $L_y(\bm{k})$ are represented as the $x$ and $y$ components of each arrow and $L_z(\bm{k})$ as the background color.  (c) Spin texture of PtSe$_{2}$ represented following the convention in (b) but replacing the orbital angular momentum with the spin operator. }
\label{fig:figure1}
\end{figure*}

{\emph{Methods}}. We performed Density Functional Theory (DFT) calculations using the plane-wave-based code \textsc{Quantum Espresso} \cite{QE-2017}. The correlation and exchange functionals are treated within the generalized gradient approximation (GGA) \cite{pbe}. We described the ionic cores using fully relativistic projected augmented wave potentials (PAW) \cite{PAW} and set an energy cutoff of $120$ Ry and $800$ Ry for computing the wave functions and charge densities, respectively. For the self-consistent run, we used a $k$-mesh of $24 \times 24 \times 1$ points, and for subsequent calculations, we set a $32\times32\times1$ mesh. To avoid spurious interactions due to periodic boundary conditions, we added a vacuum spacing of $17.57$ $\AA$ and set the forces and the self-consistent cutoffs to $5.142$ meV/$\AA$ and $4.03\times 10^{-9}$ Ry, respectively. We obtained an energy gap of 1.17 eV and a lattice constant of 3.75 $\AA$ in good agreement with previous first-principles calculations. We then constructed an effective tight-binding Hamiltonians from our DFT calculations using the pseudo-atomic orbital projection (PAO) method \cite{PAO1,PAO2,PAO3,PAO4} implemented in \textsc{PAOFLOW} \cite{PAO5,PAO6}. This approach consists in projecting the Kohn-Sham energy states into the compact subspace spanned by the pseudo-atomic orbitals built in the PAW potentials. We used PAW potentials for the Pt and Se constructed with an $spd$ and $sp$ basis, respectively.  The computation of the spin-orbit torque is achieved using $\bm{T}=\sum_i \hat{\bm{m}}\Delta_{xc}\times\langle\bm{S}^i\rangle$, where $\hat{\bm{m}}$ is the magnetization direction, $\Delta_{xc}$ is the exchange coupling strength and $\langle\bm{S}^i\rangle$ is the nonequilibrium spin density at the $i$-th atomic plane calculated with the Kubo-Bastin formula \cite{bastin1971quantum}:

\begin{multline}
     \langle\bm{S}^i(\varepsilon)\rangle = -2\text{Im}\int_{-\infty}^{\varepsilon}d\varepsilon^{\prime}f(\varepsilon^{\prime})\times \\ \left(\text{Tr}\langle \bm{S}^i\delta(\hat{H}-\varepsilon^{\prime})\partial_{\varepsilon^{\prime}}{G_{\varepsilon^{\prime}}}^{+}(\hat{\bm{J}}\cdot\bm{E}) \rangle\right),
\end{multline}

where $\bm{S}^i$ is the vector of spin density operators at the $i$-th atomic plane, $\bm{E}$ is the electric field, $\hat{\bm{J}}$ is the current density operator  $\hat{\bm{J}}=-\frac{ei}{\hbar}\left[\hat{H},\bm{R}\right]$ with $\hat{\bm{R}}$ the position operator and $\hat{H}$ the Hamiltonian of the system, and  $G_{\varepsilon}^{+}$ and $\delta(H-\varepsilon)$ the retarded Green's and spectral functions are approximated using the kernel-polynomial method \cite{kite,fan2021linear} as implemented in the \textsc{LSQUANT} toolkit \cite{lsquant_page}. We used $1024$ Chebyshev moments with the Jackson kernel to obtain an energy resolution of $\delta\varepsilon=33$ meV and fully exploit the symmetries of the system, which allows considering sizes of $256\times256$ unit cells with $34$ orbitals each, amounting to over $2$ millions of atomic orbitals with $\sim 3.067\times10^{9}$ hopping terms per unit cell achieving DFT-level precision (See SM \cite{SuplementaryMaterial}).

\begin{figure*}[t]
\includegraphics[width=0.85\linewidth]{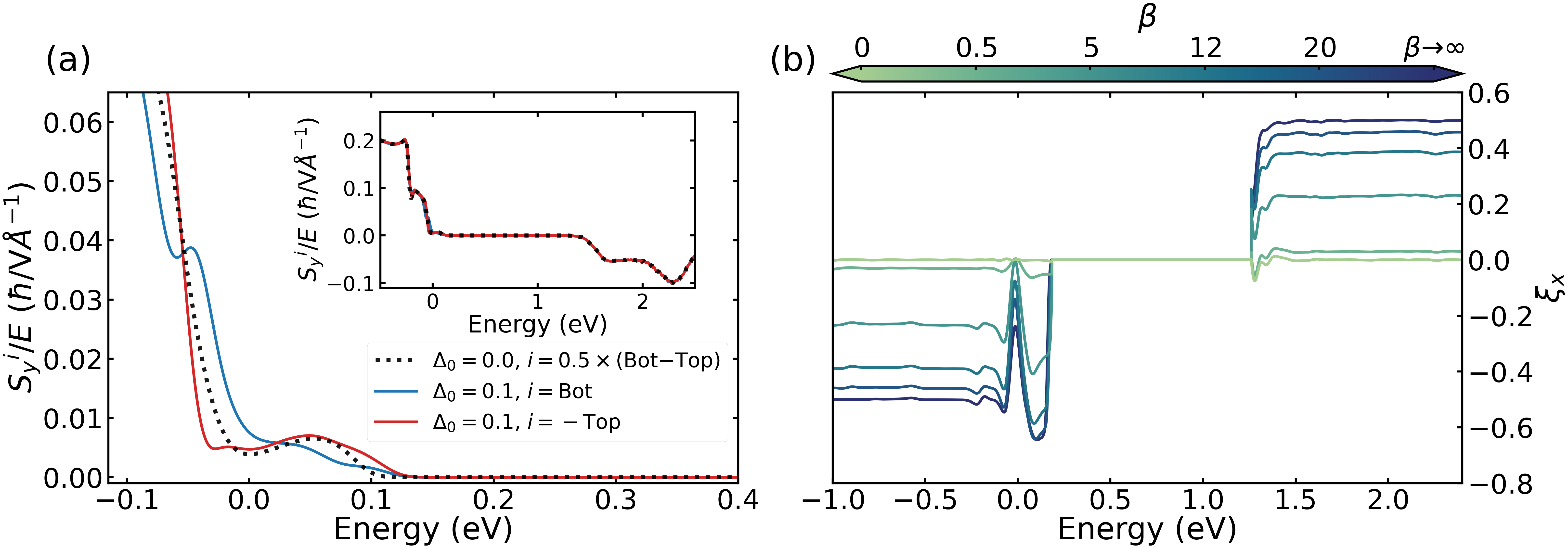}
\caption{(a) Layer-projected nonequilibrium $S_y$ densities as a function of the energy for the cases with $\Delta_0=0$ (dotted line) and $\Delta_0=100\text{ meV}/\hbar$ (solid lines) with $\beta\to\infty$. The colors indicate the $i=$top (red) and $i=$bottom (blue) contributions for the case with $\Delta_0\neq0$. Inset: Layer-projected nonequilibrium spin densities for a wider energy range. (b) Torque efficiency $\xi_x$ for $\Delta_0=100\text{ meV}/\hbar$ as a function of the energy. The colors signal the speed of the decay of the effective exchange interaction with the vertical distance.}
\label{fig:figure2}
\end{figure*}

{\it Orbital and spin-momentum locking.} \subfig{\ref{fig:figure1}}{a} shows a perfect agreement between the DFT energy bands (red dots) against those computed using the PAO Hamiltonian (blue solid lines) within the considered energy range. In the inset of \subfig{\ref{fig:figure1}}{a}, we show the crystalline structure of PtSe${}_2$ (left) together with the electric dipole field (right) computed directly from the DFT as the partial derivative of the planar average of the potential energy versus the vertical distance within the simulation box. We found a finite and sizable field located at both Se atomic planes (which are inversion partners) with opposite values. Moreover, the fields vanish on the Pt ion as a consequence of the inversion symmetry of the system. Panels \subfig{\ref{fig:figure1}}{b-c} present the orbital and spin textures nearby the Fermi contour defined by the energy $E_F=-0.05$ eV, respectively. The orbital (spin) textures are defined as the Fermi surface average of the orbital (spin) operators ($\langle\mathcal{O}_n(\bm{k})\rangle=-4K_BT\sum_n {f^{\prime}_{n\bm{k}}}\langle{\psi_{\bm{k}}}^{n}|\mathcal{O}|{\psi_{\bm{k}}}^{n}\rangle$), where $K_B$ is the Boltzmann constant, $T$ is the temperature, $f^{\prime}_{n\bm{k}}$ is the derivative of the Fermi-Dirac distribution and $|\psi_{\bm{k}}^{n}\rangle$ is the eigenstate of the $n$-th band evaluated at $\bm{k}$. The arrows in these figures indicate the $x$ and $y$ components of the orbital (spin) textures while the color represents their out-of-plane components as quantified by the color bars. \subfig{\ref{fig:figure1}}{b}
shows that within our energy range, the bands display orbital-momentum locking. The orbital textures are reversed at opposite Se planes (see SM \cite{SuplementaryMaterial}), which is related to the internal symmetries of the system.  Indeed, $1T$ monolayers of PtSe${}_2$ belong to the $D_{3d}$ point group so that the chalcogen sublayers are rotated $180^{\circ}$ to each other, yielding an equal distribution of charges across the Se-Pt-Se atomic layers. Thus, this charge distribution and the absence of a horizontal mirror plane allow the existence of the vertical dipole fields (inset of \subfig{\ref{fig:figure1}}{a}), which are similar to the ones observed in layers of bulk HfS${}_2$ \cite{clark2022hidden}. These fields induce a local out-of-the-plane asymmetry and separate the manifold of $p$ orbitals of the chalcogen atoms in two irreducible representations (Irrep). One of these representations is one-dimensional, and it is related to the $p_z$ orbitals, which are moved to lower energies, whereas the other is a $2\times2$ Irrep spanned by linear combinations of the $p_x$ and $p_y$ orbitals and are the main constituents of the top valence bands at $\Gamma$. Since the latter Irrep is $2\times2$, in close analogy with spin, the top valence bands near the Brillouin zone are mappable into orbital angular momentum (OAM) pseudospinors. In this picture, it becomes clear that the hybridizations with the $p_z$ orbitals occurring far from $\Gamma$ are translated as an effective coupling between these OAM pseudospinors, which consequently leads to the emergence of the orbital Rashba effect \cite{OrbitalRashbaDetect,Orbital-Rashba}. This mechanism explaining the appearance of orbital textures is similar to the one presented in Ref.[\onlinecite{Han2022}]. However, here the global inversion symmetry in the system imposes that the orbital (spin) textures on the two atomic chalcogen planes are oppositely aligned and degenerate in reciprocal space, leading to the vanishing of the absolute spin and OAM. Such type of orbital textures and their real-space configuration can therefore form even in the absence of spin-orbit coupling ({See SM \cite{SuplementaryMaterial}}), reinforcing their crystal field origin. Finally, \subfig{\ref{fig:figure1}}{c} shows that when the spin-orbit coupling is present, such orbital textures are imprinted into the spin features, inducing spin-momentum locking, which has the particularity to be opposite at each chalcogen atomic plane. This is in agreement with recent experimental findings by Refs. [\onlinecite{PtSe2Spin-Layer}] and [\onlinecite{clark2022hidden}], which have revealed the presence of hidden spin-textures at the surfaces of PtSe${}_2$ and HfS${}_2$. 

{\it{Atomic-plane localized spin-orbit torque}}. The dominant mechanism for SOT in conventional 2D materials is the REE. It relies on the RE that emerges in systems with broken inversion symmetry, typically due to the presence of an interface, which splits the energy bands of opposite chirality and generates a nonequilibrium spin density due to the differences in the Fermi contour. Here for PtSe${}_2$, we demonstrate that spin-momentum locking emerges naturally from the interplay between the SOC and the crystal fields. However, momentum-space degeneracy enforced by the inversion symmetry of the system prevents the formation of a net nonequilibrium spin densities. Nonetheless, practical 2D SOT devices will be typically based on PtSe${}_2$/Ferromagnetic (FM) heterostructures. Such stacking produces two effects at the origin of SOT, namely (I) a built-in electric field induced by the presence of the FM and the substrate, and responsible for the spin-splitting (Stark effect); and (II) an asymmetric electronic coupling between the closest and farthest Se-planes with respect to the FM. Generally, (I) is expected to dominate most of the conventional SOT phenomenology in \emph{centrosymmetric} systems. However, the complex crystalline structure and the real-space localization of orbital and spin textures observed in 1T PtSe${}_2$ and other 1T TMDs suggest that the asymmetric coupling between the chalcogen sublayers and the FM will disrupt the compensation of the angular momentum textures, resulting in sizable-spin-to-charge conversion and favoring the appearance of new SOT components. To fully quantify the resulting contribution of (II), driven by such localized nature of the interaction between the exchange coupling field and the orbital and spin textures in this system, we included some phenomenological position-dependent exchange term in our Hamiltonian as:

\begin{figure*}[htpb]
\includegraphics[width=0.9\linewidth]{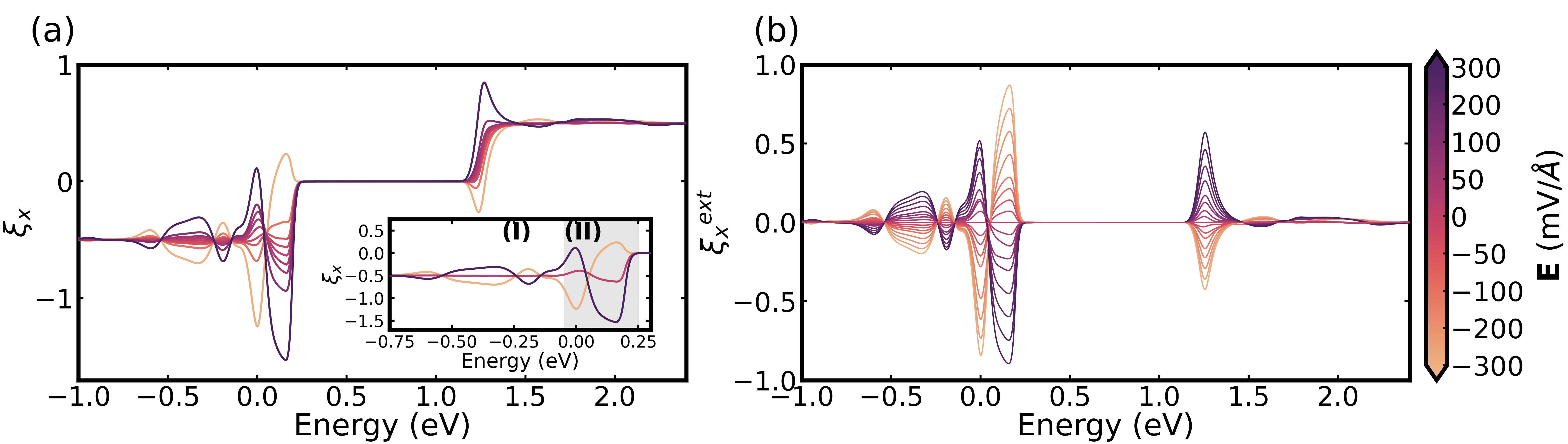}
\caption{Total and extrinsic torque efficiencies $\xi_x$ and ${\xi_x}^{ext}$, as a function of the energy for the case with $\Delta_0=100$ meV$/\hbar$ and $\beta\to\infty$ for negative and positive fields. Inset: Comparison between the torque efficiencies in the absence of external electric fields and with $\bm{E}=\pm300$ mV/$\AA$, the labels refer to the dominant contribution for SOT. The colors indicate the value of the out-of-plane electric field.}\label{fig:figure3}
\end{figure*}

\begin{equation}
    H_{xc}(\beta) = \sum_i\Delta_{0}\exp\left(-\beta\left(\frac{z_i-z_{top}}{z_{bot}}\right)^{2}\right) \cdot {S_z^{i}},
\end{equation}

where $S_z^{i}$ is the $z$ component of the spin operator of the $i$-atomic plane, $\Delta_{0}=100\text{ meV}/\hbar$ is the intensity of the exchange field, $\beta$ is a phenomenological decay factor of the proximity-induced exchange interaction, $z_{top}$ and $z_{bot}$ are the positions of the top and bottom Se atomic planes, respectively. 

We first analyze the effect associated with the contribution (II) on the spin-to-charge conversion and on the torque generation in the system. \subfig{\ref{fig:figure2}}{a} compares the energy-resolved nonequilibrium $S_y$ spin density projected at both chalcogen planes in the cases with (considering $\beta \to \infty$) and without exchange coupling. From the figure, it is clear that the coupling term breaks the inversion symmetry and disrupts the compensation of the nonequilibrium spin densities of the top and bottom Se planes. Following this, a layer-dependent energy shift occurs near the band gap and signals the real-space decoupling of the energy states of both layers in this energy region. \subfig{\ref{fig:figure2}}{a} inset, present the overall layer-dependent nonequilibrium $S_y$ spin densities for both cases. Qualitatively, the spin-to-charge conversion in this system is not altered by the inclusion of the exchange term at the top Se-plane, indicating that the resilience of these layer-localized spin-orbital textures is due to the dominating energy scale set by the crystal field. Due to the concealed nature of these spin textures and their Fermi-surface dependence, we define a specific figure of merit; the torque efficiency as $\mathbf{\xi}(\beta)=\sum_i\Delta_{xc}(\beta)\times \mathbf{S}^i(\beta)/\Delta_0 \sum_i |\mathbf{S}^{i}(\beta)|$, which quantifies the fraction of the total spin angular momentum that participates in the torque generation irrespective of the specific details associated to the bandstructure of the system.

Fig. \ref{fig:figure2}(b) shows $\xi_x$ associated with the torque component along the x-direction. One observes that $\xi_x$ increases with increasing $\beta$, reaching values as large as $\xi_x\simeq 0.5$ over most of the considered energy range. This enhancement of $\xi_x$ is related to the suppression of the coupling between the bottom Se-plane and the FM, which leaves the top Se-plane as the sole contributor to the torque intensity. In the vicinity of the charge neutrality point, $\xi_x$ exceeds $0.6$, indicating the predominance of the contribution coming from the top Se plane to the total current-induced nonequilibrium spin densities. Additionally, near the charge neutrality point, $\xi_x$ exhibits a dip that occurs for all the values of $\beta$. This reduction is related to the predominance of contributions to the spin-to-charge conversion coming from the bottom Se and Pt atoms. In contrast, the $y$-component of the  $\xi_y$ (shown in SM \cite{SuplementaryMaterial}) is only sizable near the band gap edge and inverts its value for large enough $\beta$, owing to the suppression of the SOT contributions from the bottom Se and Pt atoms.

\emph{Electrical tailoring of spin-orbit torque.} We further include the effect of an external electric field in our simulations to deepen the analysis of the interplay between the contributions (I) and (II) to the spin-orbit torque components. \subfig{\ref{fig:figure3}}{a} shows the $\xi_x$ torque efficiency for various negative and positive out-of-the-plane electric fields for large $\beta$. To facilitate the comparison with the efficiencies portrayed in Fig. \ref{fig:figure2}, we compute the torque efficiency using the total spin angular momentum in the system for the case without external electric fields as a reference to quantify the additional contributions to the torque coming from (I). Upon inspection, one can identify two regions where the effects operate differently. Away from the energy gap (shown as (I) in the inset), both field configurations produce similar changes in the torque efficiency but with opposite signs. This sign dependence shows that contributions (I) dominate the torque generation in this region. Conversely, near the charge neutrality point (shaded area labeled as (II) on the inset), the combined effect of (I) and (II) modulates the total torque efficiency. \subfig{\ref{fig:figure3}}{b} illustrates this more clearly. Here we defined the extrinsic efficiency as ${\xi_x}^{ext}=\xi_x(E) - \xi_x(E=0)$, which captures solely the effects associated with the external fields. This figure shows that for negative fields, (I) and (II) have opposite signs, in the limit of large negative fields, (I) dominates and defines the sign of the torques in the system. In contrast, for positive fields, (I) and (II) work cooperatively, and for large field amplitudes, the torque efficiency reaches approximately $\sim \times 3$ the efficiency without an external field. Thus, this electrical tunability creates opportunities for electrical control of SOT in compact devices.   

In conclusion, by combining first-principles calculations, tight-binding models, and large-scale quantum transport simulations, we uncovered the orbital origin of the hidden spin textures in inversion symmetric TMD and analyzed their real-space characteristics. Using PtSe${}_2$ as a prototypical member of the 1T TMD family, we demonstrated that the overall efficiency in generating spin-orbit torques in these systems is substantially increased by the real-space proximity interactions breaking inversion symmetry between the two chalcogen atoms (which are inversion partners). Our results suggest that other members of the 1T TMD family, such as HfS${}_2$, PtTe${}_2$ and PdSe${}_2$ could be good candidates for exploiting hidden spin-orbital textures and their real-space localization in torque applications. This discovered phenomenon, evidencing the importance of the cooperative interplay between the orbital and spin degrees of freedom, paves the way to more systematic real-space control of spin-orbit torques for developing novel spin-based technologies.

\begin{acknowledgments}
 We acknowledge discussions with A.I. Figueroa, J.F. Sierra, S.O. Valenzuela, and T.G. Rappoport. L.M.C. acknowledges funding from Ministerio de Ciencia e Innovación de Espa\~na under grant No. PID2019-106684GB-I00 / AEI / 10.13039/501100011033, FJC2021-047300-I, financiado por MCIN/AEI/10.13039/501100011033 and the European Union “NextGenerationEU”/PRTR.". J.H.G. acknowledge funding from  the European Union (ERC, AI4SPIN, 101078370). S.R and J.H.G, acknowledge funding from MCIN/AEI /10.13039/501100011033 and European Union "NextGenerationEU/PRTR”  under grant PCI2021-122035-2A-2a and funding from the European Union’s Horizon 2020 research and innovation programme under grant agreement No 881603. ICN2 is funded by the CERCA Programme/Generalitat de Catalunya and supported by the Severo Ochoa Centres of Excellence programme, Grant CEX2021-001214-S, funded by MCIN/AEI/10.13039.501100011033.
\end{acknowledgments}


%

\end{document}